\documentclass[acp, manuscript]{copernicus}

\usepackage{amsmath}

\usepackage{tikz}
\usetikzlibrary{patterns}
\usetikzlibrary{shapes}
\usetikzlibrary{decorations.text}
\usetikzlibrary{shapes.multipart}
\usetikzlibrary{arrows,shapes.geometric,positioning}
\usetikzlibrary{shapes,positioning,decorations.pathmorphing}
\usetikzlibrary{calc}
\usepackage{booktabs}
\usetikzlibrary{backgrounds}
\usepackage{varwidth}
\usetikzlibrary{fit}

\usepackage{latexsym,euscript,textcomp}
\usepackage{color,graphics,epsf,dcolumn}
\usepackage{epstopdf,ulem}
\usepackage{blindtext}
\usepackage{threeparttable}
\newcommand{\beq}{\begin{equation}}
\newcommand{\eeq}{\end{equation}}

\begin{document}
\nolinenumbers

\title{On the Methodology for Assessing Vegetation Impacts on the Atmospheric Branch of the Hydrological Cycle}


\Author[1,2]{A. M.}{Makarieva}
\Author[1,2]{A. V.}{Nefiodov}
\Author[2]{A. D.}{Nobre}
\Author[3,2]{L.A.}{Cuartas}
\Author[2,4]{F.}{Pasini}
\Author[2,4]{D.}{Andrade}

\affil[1]{Theoretical Physics Division, Petersburg Nuclear Physics Institute of National Research Center {\textquotedblleft}Kurchatov Institute{\textquotedblright}, St.~Petersburg, Russia}
\affil[2]{Biotic Pump Greening Group Institute, S\~{a}o Jos\'{e} dos Campos, Brazil}
\affil[3]{General Coordination of Research and Development, National Center for Monitoring and Early Warning of Natural Disasters (CEMADEN), S\~{a}o Jos\'{e} dos Campos, Brazil}
\affil[4]{Federal University of Rio de Janeiro, Rio de Janeiro, Brazil}


\runningtitle{On the Methodology for Assessing Vegetation Impacts}

\runningauthor{A. M. Makarieva et al.}

\correspondence{Anastassia Makarieva (ammakarieva@gmail.com)}

\received{}
\pubdiscuss{} 
\revised{}
\accepted{}
\published{}


\firstpage{1}

\maketitle

\begin{abstract}
\large
China has undertaken unprecedented, state-driven vegetation restoration on a continental scale. This large-scale land-surface intervention offers a rare opportunity to assess how deliberate biospheric change influences climate-relevant processes, especially the hydrological cycle. Of particular interest is how increased water use by additional vegetation affects terrestrial water availability, including streamflow that sustains both ecosystems and human society.

Here we evaluate the methodological basis for addressing this question in light of recently available data on hydrological change in China. Revisiting the atmospheric branch of the hydrological cycle, we argue that water yield depends fundamentally on vegetation-induced changes in atmospheric circulation. When the effects of vegetation on atmospheric dynamics are neglected, as in moisture-recycling-based approaches, the analysis is predisposed by construction toward diagnosing a negative effect of additional vegetation on water yield.

Given the nonlinear dependence of precipitation on atmospheric moisture, we further suggest that streamflow reductions associated with added vegetation in dry regions may reflect a transient phase of early ecological succession rather than a persistent long-term outcome. As ecosystems mature and regional moisture regimes evolve, this relationship may reverse, generating a positive feedback between vegetation cover and water availability. We briefly discuss recent observational evidence consistent with this interpretation.

We conclude that robust assessment of vegetation impacts on water yield requires frameworks that explicitly couple vegetation change, atmospheric processes, and hydrological responses. Such an approach is essential for distinguishing short-term trade-offs from longer-term system trajectories and for informing sustainable land management under continued ecosystem restoration and conservation.
\end{abstract}


\large
\section{\normalsize Introduction: Water yield, atmospheric circulation, and moisture recycling}
\label{intr}

China{\textquoteright}s large-scale re-greening efforts, implemented as part of a broader strategy to build an Eco-Ci\-vi\-li\-za\-tion \citep{Liu2025}, represent one of the most extensive land-surface interventions of recent decades. These efforts provide a unique opportunity to examine how deliberate changes in vegetation cover influence the hydrological cycle. While recent studies have made important progress in quantifying runoff and streamflow responses to re-greening, differences in methodological choices---particularly in the treatment of atmospheric processes---can lead to divergent interpretations of vegetation impacts on water yield. Clarifying the implications and limitations of these approaches is therefore essential.

Water yield is defined as the difference between precipitation \(P\) and evapotranspiration \(E\):
\begin{equation}\label{Y}
Y = P - E.
\end{equation}
It is equal to atmospheric moisture convergence, that is, the net amount of moisture delivered to a given location by the atmosphere. Water yield is partitioned into surface runoff, subsurface runoff, and, under non-steady-state conditions, recharge of soil moisture and groundwater.

A recent assessment of vegetation impacts on China{\textquoteright}s water cycle by \citet{An2025} concluded that recent land-use and land-cover change (LUCC) has reduced China{\textquoteright}s water yield. Their analysis estimates LUCC-induced changes in evapotranspiration using prescribed characteristic values for different vegetation types at each time step. For example, when grassland is converted to forest, the LUCC-induced change in evapotranspiration is defined as the difference between the actual evapotranspiration of forest and the hypothetical evapotranspiration of grassland under otherwise identical local conditions. The associated LUCC-induced change in precipitation is then estimated by tracking how much of this additional evapotranspiration re-precipitates within China using a moisture-tracking model.

This constitutes the moisture-recycling approach, in which water vapor is treated as a passive tracer transported by a prescribed atmospheric circulation. Within this framework, the impact of LUCC is reduced to determining what fraction of precipitated water is returned to the atmosphere through eva\-po\-trans\-pi\-ra\-tion.

Suppose that added vegetation increases evapotranspiration $E$ by $\Delta E_r >0$. By definition, the cor\-res\-pon\-ding change in precipitation $P$ due to recycling of this additional evapotranspiration, $\Delta P_r$, cannot exceed the added evapotranspiration itself, because part of the evaporated moisture may leave the region without precipitating; thus, 
\begin{equation}\label{PE}
\Delta P_r \le \Delta E_r.
\end{equation}
Consequently, when defined in this way, the LUCC-induced change in water yield $Y$ associated with moisture recycling,
\begin{equation}\label{Yr}
\Delta Y_r = \Delta P_r - \Delta E_r \le 0,
\end{equation}
is negative by construction for any positive $\Delta E_r \ge 0$.

\begin{figure*}[tb]
\begin{minipage}[p]{0.9\textwidth}
\centerline{\includegraphics[width=1\textwidth,angle=0,clip]{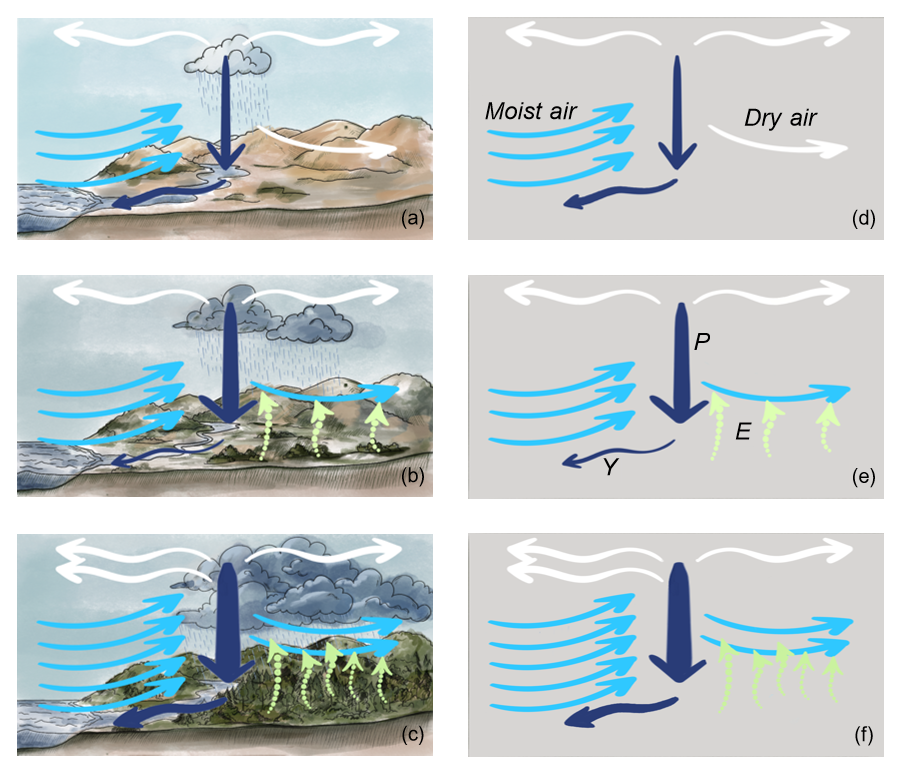}}
\end{minipage}
\caption{\normalsize Different effects of added transpiration on water yield. Light-blue and white horizontal arrows denote moist and dry air, respectively; the dark-blue vertical arrow denotes precipitation; and the dark-blue horizontal arrow denotes water yield (for simplicity, shown as river runoff). Wavy light-green arrows denote evapotranspiration. The panels in the left and right columns depict the same air circulation and hydrological cycle, but in the right column the landscape is omitted for clarity. Panels (a) and (d) show the hypothetical case of zero evapotranspiration. Panels (b) and (e) show the effect of moisture recycling alone: the air circulation is the same as in panels (a) and (d), but the region becomes a source of moist air, while local water yield is reduced accordingly relative to panels (a) and (d). Panels (c) and (f) show increased moisture convergence and water yield due to added evapotranspiration (the biotic pump regime). }
\label{fig1}
\end{figure*}

To illustrate this point, consider a simplified land region bordering the ocean (Fig.~\ref{fig1}). Winds transport moist air inland, where, in the absence of evapotranspiration, atmospheric moisture is removed by pre\-ci\-pi\-ta\-tion and partially dried air exits the region. In this limiting case, all precipitation contributes directly to runoff (Fig.~\ref{fig1}a,d).

If evapotranspiration is introduced while neither the circulation nor the moisture content of the incoming oceanic air changes, part of the precipitated water is returned to the atmosphere, increasing the moisture content of the air leaving the region. This reduces moisture convergence, defined as the difference between gross atmospheric moisture inflow and outflow. Thus, as required by mass conservation, moisture recycling alone implies a decrease in local water yield (Fig.~\ref{fig1}b,e).

An increase in moisture convergence---and thus in runoff and water yield---is only possible if atmospheric circulation changes. Specifically, moisture-laden air must either enter the region more rapidly or leave it more slowly (Fig.~\ref{fig1}c,f; see \citet{Makarieva2025front} for a detailed discussion).

Vegetation modifies key land-surface properties, including albedo and aerodynamic roughness \citep{yang19,breil2021}, and---through evapotranspiration---affects near-surface tem\-pe\-ra\-tu\-re \citep{Barnes2024}, atmospheric humidity \citep{jiang13}, and cloud formation \citep{Lyons2002}. These processes can, in turn, alter atmospheric circulation and moisture transport.  The enhancement of moisture convergence due to added forest evapotranspiration constitutes the essence of the biotic pump \citep{hess07,makarieva23}.

By excluding these pathways and reducing the impact of land-use and land-cover change solely to moisture recycling, as in Fig.~\ref{fig1}b,e, one imposes a strong \emph{a priori} constraint on the interpretation of vegetation effects on the water cycle. Under this formulation, any re-greening region with increasing evapotranspiration is, by construction, diagnosed as having a vegetation-induced decline in local water yield given by Eq.~(\ref{Yr}), irrespective of circulation-mediated responses or actual water-yield trends.

\section{\normalsize Model and data challenges}

\begin{table}[h]
\centering
\begin{threeparttable}
\caption{Reported annual rates of change in evapotranspiration, precipitation, and water yield  
in China during the Grain For Green Program from different sources and methods. Units: mm\,year$^{-2}$.}
\label{tab1}

\begin{tabular}{c c c l l }
\toprule
{$\Delta E$} & {$\Delta P$} & {$\Delta Y$} & Region& Estimates/Data \\
\midrule
 1.71 & 1.24 & $-0.46$ & China\tnote{a} & Estimated LUCC-induced changes, Eq.~(\ref{Yr}) \\
$-0.86$ & 0.35 &  1.21 & China\tnote{b} & ERA5 data \\
1.3 & 3.1 & 1.8 & Loess Plateau\tnote{c} & MERRA-2 data\\
\bottomrule
\end{tabular}

\begin{tablenotes}
\footnotesize
\item[a] \citet{An2025}, their Fig.~4 b,d,f: Mean annual $\Delta P_r$, $\Delta E_r$ and $\Delta Y_r$ in $2001$--$2020$ 
\item[b] \citet{An2025}, their Fig.~5a:  Mean annual $\Delta P$, $\Delta E$ and $\Delta Y$ in $2001$--$2020$ 
\item[c] \citet{tian2022}, their Fig.~10: Rainy season data, mean annual $\Delta P$, $\Delta E$ and $\Delta Y$  between $1982$--$1998$ and $1999$--$2018$ 
\end{tablenotes}

\end{threeparttable}
\end{table}

How can local and regional vegetation impacts on atmospheric circulation be disentangled, in principle, from external influences such as global climate change? Because global climate models aim to provide comprehensive representations of the Earth system, a straightforward approach is to compare simulations with and without land-use and land-cover change. However, current models do not yet demonstrate robust skill in representing vegetation-induced changes in atmospheric moisture convergence. Modeled responses to large-scale vegetation change differ widely---not only in magnitude but even in sign---across models \citep{Luo2022}. Accordingly, ensemble approaches generally fail to yield a statistically robust vegetation signal in water-yield responses \citep{Gudmundsson2021}.

Hydrological data pose a closely related challenge. Evapotranspiration cannot be measured directly and must be inferred from other observations (e.g., flux-tower measurements) using multiple parameterizations. The most robust estimates derive from watershed-scale mass balance, calculated as the difference between measured precipitation and water yield \citep{Teuling2018}. Such estimates are inherently non-local and rely on streamflow observations that remain uncertain in many regions and have only recently become available at national scale in China \citep{Wang2025}. Consequently, in reanalysis products eva\-po\-trans\-pi\-ra\-tion is a parameterized rather than observed quantity.

In ERA5 and ERA5-Land, evapotranspiration is a land-surface model diagnostic that is not independently constrained by observations, but derived from parameterized soil--vegetation--atmosphere exchanges \citep{MunozSabater2021}. This limitation is highlighted by the mismatch between the LUCC-induced upward trend in evapotranspiration over China reported by \citet{An2025} and the downward trend in actual evapotranspiration diagnosed directly from ERA5 for the same period [Table~\ref{tab1}; Fig.~5a in \citet{An2025}]. Although the LUCC-induced trend was described as robust, it likely reflects changes in vegetation cover---that is, transitions between parameterizations---rather than independently constrained changes in actual evapotranspiration. The resulting discrepancy between the two trends (Table~\ref{tab1}) remains unresolved. In particular, the apparent decline in evapotranspiration despite substantial re-greening suggests that the underlying processes are not yet consistently represented and warrant further investigation.

A plausible interpretation is that neither evapotranspiration trend is robust, because the underlying trends in precipitation and water yield---quantities that are, in principle, observable---are themselves insufficiently constrained. Consistent with ERA5, observations indicate increasing precipitation across many regions of China \citep{Tian2022warmer,Zhang2022hum,Zhan2023}, as well as a robust increase in water yield in the south \citep{Zhang2023wy}. However, their magnitudes remain uncertain, preventing reliable inference of long-term evapotranspiration change. Table~\ref{tab1} also shows data for the rainy season of the Loess Plateau \citep{tian2022}, where increased eva\-po\-trans\-pi\-ra\-tion is associated with increased moisture convergence. When hydrometeorological trends change sign across regions, seasons, and timescales, robust observational constraints are required for reliable quantification and strategic plan\-ning.

\section{\normalsize  Toward a cross-disciplinary synthesis: Combining insights from atmospheric dynamics and ecology}

While data continue to accumulate to inform improved models, there have been calls for broader use of theoretical concepts that, although less detailed than numerical simulations, can clarify the larger physical picture and guide interpretation \citep{Byrne24}.

Recent global forest-change assessments indicate that the dominant hotspots of forest loss and dis\-tur\-bance during 2000–2020 were tropical forests in the Amazon Basin, the Congo Basin, and Southeast Asia, together with extensive tree-cover loss and disturbance in boreal regions of Canada and Russia \citep{Hansen2013,Tyukavina2022,FAO2020}. In contrast, several temperate and subtropical regions---notably parts of Asia---have exhibited net forest and vegetation gain over the same period, driven in particular by large-scale re-greening in China and India \citep{Chen2019}.

Consistent with this pattern, global maps of water-yield trends for 2000--2020 highlight tropical regions, Canada, and Siberia as areas with a tendency toward declining water yield, while India and China stand out as among the few regions where water yield has increased, at least at regional scales [see Fig.~1 of \citet{Zhang2023wy}]. This spatial correspondence suggests a positive coupling between vegetation change and water yield, adding to empirical evidence for positive relationships between evapotranspiration and atmospheric moisture convergence \citep{ch10,Wright2017}.

\begin{figure*}[tb]
\begin{minipage}[p]{0.9\textwidth}
\centerline{\includegraphics[width=1\textwidth,angle=0,clip]{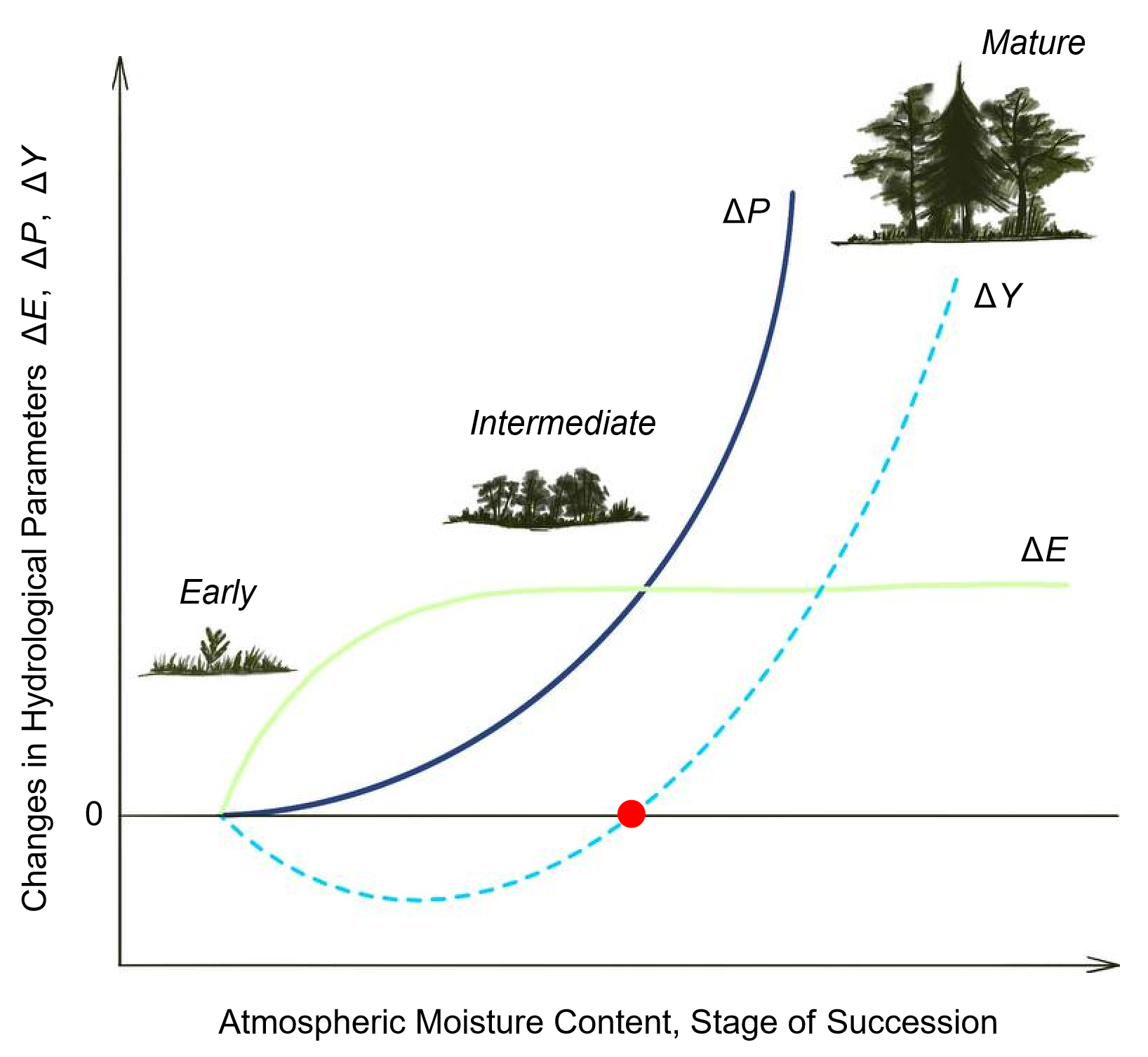}}
\end{minipage}
\caption{\normalsize Conceptual dependence between atmospheric moisture content, stage of ecological succession and changes in hydrological parameters---precipitation $P$, evapotranspiration $E$ and water  yield $Y$---relative to their initial values.  The red dot indicates the transitional point when the water yield trend becomes positive.
}
\label{fig2}
\end{figure*}

It has been proposed, based on the distinct relationships between transpiration and atmospheric moisture content, on the one hand, and precipitation and atmospheric moisture content, on the other, that ecological restoration may proceed through two hydrological stages \citep{makarieva23}. In the early stage of succession, vegetation is sparse and transpiration increases with greenness, but the atmosphere remains relatively dry and precipitation responds weakly to increasing moisture (Fig.~\ref{fig2}). In this regime, transpiration increases more rapidly than precipitation, while water yield declines, $\Delta Y < 0$. 

As succession progresses, sustained transpiration moistens the atmosphere further, and beyond a thres\-hold precipitation may respond more strongly to increasing atmospheric moisture content (Fig.~\ref{fig2}). At that point, water yield can increase with vegetation and atmospheric moisture content, $\Delta Y > 0$, consistent with a regime in which enhanced evapotranspiration is associated with increased moisture convergence (the biotic pump regime). Additional moisture convergence further reinforces atmospheric moistening, while transpiration may moderate under moist conditions as vapor pressure deficit declines.

This dual hydrological behavior during ecological restoration parallels the ecological concept of a landscape trap \citep{lindenmayer22}. Mature, relatively undisturbed forest ecosystems tend to be hydrologically stable and resistant to droughts and fires 
\citep{xiao2023land,wolf2023canopy}. Logging or other disturbances can shift such systems toward early successional states that are less hydrologically competent. If disturbance exceeds a threshold, the system may enter a trajectory of drying and burning from which recovery becomes increasingly difficult. The inability of early successional systems to stabilize a moist regime is consistent with the characteristics of the dry hydrological stage as shown in Fig.~\ref{fig2}.

In China, contrasting assessments of the relationship between water yield and ecological restoration reflect the complexity of this transition. After two to three decades of re-greening, large areas remain in relatively early successional stages, often dominated by naturally recovering grasslands rather than mature forests \citep{Yu2023}. In such conditions, analyses confined to moisture recycling alone may suggest that further ecological restoration threatens local streamflow. However, if restoration is viewed as a dynamic progression between hydrological regimes, short-term reductions in water yield need not represent a long-term trajectory.

Moisture recycling is a key pathway through which vegetation interacts with atmospheric dynamics. Understanding not only its immediate mass-balance implications but also its role in modifying atmospheric moisture convergence is essential for designing robust strategies that align ecosystem recovery with hydrological resilience. Integrating ecological succession theory with atmospheric dynamics may therefore provide a more comprehensive framework for anticipating the long-term evolution of water cycles in re-greening regions.

\citet{An2025} propose that a {\textquoteleft}smart afforestation{\textquoteright} is required that prioritizes deep-rooted, drought-tolerant species and multispecies mixtures to enhance resilience to pests and climate change. We would add that this should be understood not simply as the selection of resistant species, but as the deliberate assembly of biodiverse, multi-layer successional systems in which species are selected for complementary functional traits, including successional status, shade tolerance, rooting depth, and hydraulic strategy \citep{andrade20,pasini2021}. Such designs can increase complementarity in light capture and resource use, sustain higher photosynthetic capture per unit land area through time, and improve resistance and resilience to drought relative to monospecific stands \citep{jacobi2025}. By distributing biomass production, litter inputs, and rhizodeposition across canopy and soil layers, these systems can also set in motion the processes of soil aggregation, organic matter stabilization, and improved infiltration capacity \citep{vaupel2025}, outcomes that are better understood as successional trajectories unfolding over decades than as designed results of planting alone. Over time, such soil development creates the structural and functional conditions for progression toward a mature, late-successional forest with increasing hyd\-ro\-lo\-gi\-cal competence. In this context, higher evapotranspiration should not be interpreted as a simple depletion of soil water. In structurally complex forests, greater evapotranspirative flux is coupled with improved water capture, storage, and redistribution within the soil profile, while also reinforcing at\-mos\-phe\-ric moisture recycling and the vegetation-atmosphere feedbacks that sustain the biotic pump. As ecosystems mature and atmospheric moisture conditions evolve, the relationship between vegetation cover and water yield may reverse. Smart afforestation should therefore be oriented not only toward tree survival under drought, but toward the establishment of forests that progressively rebuild soil function and restore hyd\-ro\-lo\-gi\-cal regulation.

Finally, the high hydrological competence of minimally disturbed ecosystems, together with the slow nature of ecological succession, provides a strong rationale for preserving intact vegetation wherever it remains. Within China, there is increasing recognition that tree cover constitutes a foundation of en\-vi\-ron\-men\-tal stability. A consistent extension of this principle is to support the protection of natural forests beyond national boundaries, particularly in Eurasia and the tropics, where deforestation exerts a disproportionate influence on atmospheric circulation and climate stability. In this context, the concept of ecological civilization can be understood not only as a national development framework, but as a basis for coordinated action to preserve one of the Earth{\textquoteright}s most effective climate-regulating systems: natural forests. Safeguarding these systems is not simply a matter of conservation, but a prerequisite for maintaining the stability of regional and global water cycles.


\end{document}